\documentclass[12pt]{article}

\usepackage[latin9]{inputenc}
\usepackage{epsfig,amsmath,amssymb}
\usepackage{color}
\usepackage{graphicx}
\usepackage{cite}  
\usepackage{epsfig,amsmath,amssymb}
\usepackage{color}
\usepackage{float}
\usepackage{color}
\usepackage{bm}
\usepackage{amsmath}
\usepackage{amssymb}
\makeatletter

\@ifundefined{textcolor}{}
{%
 \definecolor{BLACK}{gray}{0}
 \definecolor{WHITE}{gray}{1}
 \definecolor{RED}{rgb}{1,0,0}
 \definecolor{GREEN}{rgb}{0,1,0}
 \definecolor{BLUE}{rgb}{0,0,1}
 \definecolor{CYAN}{cmyk}{1,0,0,0}
 \definecolor{MAGENTA}{cmyk}{0,1,0,0}
 \definecolor{YELLOW}{cmyk}{0,0,1,0}
}

\usepackage{latexsym}
\usepackage[ngerman, english]{babel}

\makeatother

\begin{document}

\author
{ P. M\"uller$^1$, J. Richter$^1$, A. Hauser$^1$ and D. Ihle$^2$\\
\small{$^1$ Institut f\"ur Theoretische Physik, Universit\"at Magdeburg, 39016 Magdeburg, Germany}\\
\small{$^{2}$ Institut f\"ur Theoretische Physik, Universit\"at Leipzig, D-04109 Leipzig,
Germany}\\}

\title{Thermodynamics of the frustrated $J_{1}$-$J_{2}$ Heisenberg ferromagnet
on the body-centered cubic lattice with arbitrary spin}

\date{\today}
\maketitle
\begin{abstract}
We use the spin-rotation-invariant  Green's function method
as well as the high-temperature expansion to discuss  the thermodynamic
properties  of the frustrated spin-$S$ $J_{1}$-$J_{2}$
Heisenberg magnet on the body-centered cubic lattice.
We consider 
ferromagnetic nearest-neighbor bonds $J_1 < 0$ and antiferromagnetic
next-nearest-neighbor bonds $J_2 \ge 0$ and arbitrary spin $S$.
We find that the transition point $J_2^c$ between the ferromagnetic ground
state and the
antiferromagnetic one is nearly independent of the spin    
$S$, i.e., it is very close to the classical transition point  $J_2^{c,{\rm clas}}=
\frac{2}{3}|J_1|$. 
At finite temperatures we focus on the parameter regime  $J_2<J_2^c$  with a ferromagnetic
ground-state.  
We calculate the Curie temperature $T_{C}(S,J_{2})$ and derive 
an empirical formula describing the influence of the frustration
parameter $J_{2}$ and spin $S$
on $T_C$. We find that the Curie temperature monotonically decreases with
increasing frustration $J_2$, where very
close to  $J_2^{c,{\rm clas}}$ the $T_C(J_2)$-curve exhibits a fast decay 
which is well described by a
logarithmic term $1/\textrm{log}(\frac{2}{3}|J_1|-J_{2})$.
To characterize the magnetic ordering below and above $T_C$,
we calculate  the spin-spin correlation
functions
$\langle {\bf S}_{\bf 0} {\bf S}_{\bf R} \rangle$, the spontaneous
magnetization, the uniform static susceptibility $\chi_0$ as well as  the correlation length
$\xi$. Moreover, we discuss the specific heat $C_V$ and the temperature
dependence of the excitation spectrum.
As  approaching the transition point $J_2^c$ some unusual features were
found, such as negative spin-spin correlations at temperatures above $T_C$ even
though the ground state is ferromagnetic or 
an increase of
the spin stiffness with growing temperature.
\end{abstract}

\section{Introduction\protect \\
}
\label{sec:intro}

Heisenberg models with competition between 
nearest-neighbor (NN) bonds $J_1$ and next-nearest-neighbor (NNN) bonds
$J_2$ 
($J_{1}$-$J_{2}$ models)  
can serve as canonical models to study frustration effects in magnetic
systems.
In case of ferromagnetic NN exchange, $J_1 <0$, and antiferromagnetic NNN
exchange, $J_2 >0$, the ferromagnetic ground
state gives way for an antiferromagnetic state with zero net magnetization at a critical value
$J_2^c$ that depends on the dimension $D$ of the
system \cite{bader,hamada,krivnov2007,shannon2006,richter2010}.
In lower dimension $D<3$ and for low spin quantum number $S$
the ground state for $J_2>J_2^c$ may have
unconventional properties, see e.g.
Refs.~\cite{krivnov2007,shannon2006,LC,momoi2011}.
Although, in the ferromagnetic regime, $J_2<J_2^c$, the ground state is
the simple fully polarized ferromagnetic state, the influence of the frustrating coupling $J_2$ on the
thermodynamic properties  can be strong, as it has been recently discussed for systems with
dimension $D=1$ and
$D=2$ \cite{selke77,tmrg,RGMchainarbitraryspin,RGMchainfrusferro,RGM2Dj1j2frusferro,dmitriev_clas,sirker2011}.      

In the present study we extend our previous studies of the   
frustrated $J_{1}$-$J_{2}$ Heisenberg ferromagnet in dimensions
$D=1$
(Refs.~\cite{RGMchainarbitraryspin,RGMchainfrusferro,sirker2011}) and
$D=2$ (Ref.~\cite{RGM2Dj1j2frusferro}) to the three-dimensional case.
In $D=3$ the  body-centered cubic (BCC) lattice is appropriate to compare
with $D=1$ and $D=2$, since the antiferromagnetic  $J_2$ bonds are in competition with
$J_1$, but do not frustrate each
other \cite{bcc_schmidt2002,bcc_oitmaa2004,bcc_majumdar2009,bcc_momoi2013,bcc_RPA_2014}.

The Hamiltonian of the spin-$S$ $J_1$-$J_2$ Heisenberg model on the BCC
lattice considered in this
paper is given by
\begin{equation}
H=J_{1}\sum_{\langle i,j\rangle}\bm{S}_{i}\cdot
\bm{S}_{j}+J_{2}\sum_{\left[i,j\right]}\bm{S}_{i}\cdot\bm{S}_{j} \; , \;
 \label{eq_ham}
\end{equation}
where $\bm{S}_{i}^{2}=S(S+1)$, and $\langle i,j\rangle$ runs over NN bonds and  $[i,j]$ over all NNN
bonds.
We consider  $J_1 < 0 $ and $J_2 \ge 0$. In what follows we define  the energy
scale by setting $J_1=-1$.
In the classical limit $S \to \infty$, the ground state is ferromagnetic for
$J_2 < J_2^{c,{\rm clas}}=\frac{2}{3}|J_1|$, and it is a collinear antiferromagnetic state with
the magnetic wave-vector ${\bf Q}=(\pi,\pi,\pi)$  for $J_2 >
J_2^{c,{\rm clas}}$ \cite{bcc_schmidt2002,bcc_oitmaa2004,bcc_majumdar2009,bcc_momoi2013}.
In what follows we want to discuss the thermodynamics of the model
(\ref{eq_ham})
for arbitrary spin quantum number $S$ and focus on the 
parameter regime $J_2<J_2^c$, where the ferromagnetic ground state is
realized.

As in our previous papers on frustrated
ferromagnets,\cite{RGMchainarbitraryspin,RGMchainfrusferro,RGM2Dj1j2frusferro} 
we use the  
rotationally invariant Green's function method (RGM)
to study the influence of the
frustrating NNN bond $J_2$ and of the spin
quantum number $S$ on the thermodynamic properties of the model  (\ref{eq_ham}).
This Green's function approach  
has been applied successfully to 
several  quantum spin systems
\cite{barabanov94,winterfeldt97,SIH00,yu_feng2000,ihle2001,canals2002,prb2004,schmal2006,RGMquasi2Dvskagomelattice,RGMferromagneticfield,antsyg,antsyg2012,heisenbergferro2d,RGMSgg1anisotropy,
RGMlayeredheisenbergarbitraryspin,RGMchainfrusferro,RGM2Dj1j2frusferro,RGMchainarbitraryspin,kondoyamaji,RhoScal94,SSI94,ShiTak1991,RGMmoritzcollinearstripe,RGMchainfrusferromagnet}.
We complement the Green's function study by 
using the high-temperature expansion (HTE) technique \cite{domb_green,OHZ06,HTE,HTE2}.

The paper is organized as follows: In Sec.~\ref{methods},
the RGM is developed for the model \eqref{eq_ham} and a brief
instruction of the HTE approach is given.
Some relevant features of the zero-temperature transition between the ferromagnetic
ground-state and the collinear antiferromagnetic state are discussed  
in Sec.~\ref{sec:GS}.
Finite-temperature properties are then analyzed in Sec.~\ref{finite_T}.
We summarize our findings in Sec.~\ref{sec:sum}.

\section{Methods}
\label{methods}

\subsection{Rotation-invariant Green's function method (RGM) \label{RGM}}
As already mentioned above, the RGM is a well established method in the field
of
magnetic systems including strongly frustrated quantum spin systems. 
It was initially  introduced by Kondo and Yamaji
for spin-$1/2$ systems \cite{kondoyamaji}. Later on, an extension to
arbitrary spin  $S$ was developed, see, e.g.,
Refs.~\cite{RhoScal94,SSI94,RGMSgg1anisotropy,RGMlayeredheisenbergarbitraryspin,RGMchainarbitraryspin}.

To determine the spin-spin correlation 
functions and the thermodynamic quantities, we 
calculate the transverse dynamic spin susceptibility $\chi_{\mathbf{q}}^{+-}(\omega)=-\langle\langle S_{\mathbf{q}}^{+};S_{-\mathbf{q}}^{-}\rangle\rangle_{\omega}$
(here, $\langle\langle\ldots;\ldots\rangle\rangle_{\omega}$ denotes
the two-time commutator Green's
function\cite{Tya67,gasser,QuantumTheoryoMagn}).
Taking the equation of motion up to the second step (the decoupling of
the higher order Green's function is performed in the second-order equation of
motion, i.e., one order beyond standard random-phase
approximation) and supposing
spin rotational symmetry, i.e., $\langle S_{i}^{z}\rangle=0$, we
obtain $\omega^{2}\langle\langle S_{\mathbf{q}}^{+};S_{-\mathbf{q}}^{-}\rangle\rangle_{\omega}=
M_{\mathbf{q}}+\langle\langle-\ddot{S}_{\mathbf{q}}^{+};S_{-\mathbf{q}}^{-}\rangle\rangle_{\omega}$
with $M_{\mathbf{q}}=\left\langle \left[[S_{\mathbf{q}}^{+},H],S_{-\mathbf{q}}^{-}\right]\right\rangle $
and $-\ddot{S}_{\mathbf{q}}^{+}=\left[[S_{\mathbf{q}}^{+},H],H\right]$.
For the model (\ref{eq_ham}) the moment $M_{\mathbf{q}}$ is given
by the exact expression 
\begin{equation}
M_{\mathbf{q}}=-2J_{1}z_{1}c_{001}(1-\gamma_{\mathbf{q}}^{(1)})-2J_{2}z_{2}c_{101}(1-\gamma_{\mathbf{q}}^{(2)}),\label{eq_mq}
\end{equation}
where $c_{hkl}\equiv c_{\bm{R}}=\langle S_{\bm 0}^{+}S_{\bm{R}}^{-}\rangle=2\langle
{\bf S}_{\bm 0}{\bf S}_{\bm{R}}\rangle/3$,
$\bm{R}=h\bm{a}_{1}+k\bm{a}_{2}+l\bm{a}_{3}$, $\bm{a}_{i}=\frac{1}{2}\sum_{j}\left[\left(1-2\delta_{j,i}\right)\bm{e}_{j}\right]$
($\bm{e}_{j}$ are the Cartesian unit vectors), $\gamma_{\mathbf{q}}^{(1)}=\cos\frac{q_{x}}{2}\cos\frac{q_{y}}{2}\cos\frac{q_{z}}{2}$
and $\gamma_{\mathbf{q}}^{(2)}=\frac{1}{3}(\cos q_{x}+\cos q_{y}+\cos q_{z})$.
The quantities $z_1=8$ and $z_2=6$ are the NN and NNN coordination numbers
of the BCC lattice,
respectively.  
For the second derivative $-\ddot{S}_{\mathbf{q}}^{+}$ we use the
standard decoupling of the RGM, see, e.g.,
Refs.~\cite{SSI94,RGMchainfrusferro,heisenbergferro2d,kondoyamaji,RGM2Dj1j2frusferro,
RGMchainarbitraryspin, RGMchainfrusferromagnet,
RGMlayeredheisenbergarbitraryspin,ShiTak1991,RhoScal94,SIH00,yu_feng2000,ihle2001,winterfeldt97,barabanov94}. To be 
specific, in $-\ddot{S}_{i}^{+}$ we apply the decoupling in real
space 
\begin{align}
S_{i}^{+}S_{j}^{+}S_{k}^{-} & =\alpha_{i,k}\langle S_{i}^{+}S_{k}^{-}\rangle S_{j}^{+}+\alpha_{j,k}\langle S_{j}^{+}S_{k}^{-}\rangle S_{i}^{+},\label{eq_entk1}\\
S_{i}^{+}S_{j}^{-}S_{j}^{+} & =\langle S_{j}^{-}S_{j}^{+}\rangle S_{i}^{+}+\lambda_{i,j}\langle S_{i}^{+}S_{j}^{-}\rangle S_{j}^{+},
\label{vertex_lambda}
\end{align}
where  $i\neq j\neq k\neq i$.
The quantities $\alpha_{i,j}$ and $\lambda_{i,j}$ are vertex parameters
introduced to improve the
decoupling approximation.
Note that 
products of three spin operators with two coinciding sites, see
Eq.~(\ref{vertex_lambda}), appear only for $S\ge 1$, i.e., 
the  vertex parameter
$\lambda_{i,j}$ is irrelevant for
$S=1/2$ \cite{SSI94,RGMlayeredheisenbergarbitraryspin,RGMchainarbitraryspin}.
By using
the operator identity $\bm{S}_{i}^{2} 
=S_{i}^{+}S_{i}^{-}-S_{i}^{z}+(S_{i}^{z})^{2}$ the expectation value $\langle
S_{j}^{-}S_{j}^{+}\rangle$ entering Eq.~(\ref{vertex_lambda}) is given by
\begin{align} \label{eq:op_ident}
\langle S_{j}^{-}S_{j}^{+}\rangle=\langle S_{j}^{+}S_{j}^{-}\rangle &
=\frac{2}{3}S(S+1),
\end{align}
where $\langle S_j^z\rangle=0$ was used.
 For systems with a ferromagnetic ground state it is known from Refs.~\cite{heisenbergferro2d,
RGM2Dj1j2frusferro, RGMchainfrusferromagnet, RGMferromagneticfield,
RGMlayeredheisenbergarbitraryspin} that we may set  $\alpha_{i,k}=\alpha$
and $\lambda_{i,j}=\lambda$, i.e., we approximate the various
multispin correlation functions appearing in the equation of motion on an
equal footing.
We obtain
$-\ddot{S}_{\mathbf{q}}^{+}=\omega_{\mathbf{q}}^{2}S_{\mathbf{q}}^{+}$
and 
\begin{equation}
\chi_{\mathbf{q}}^{+-}(\omega)=-\langle\langle S_{\mathbf{q}}^{+};
S_{-\mathbf{q}}^{-}\rangle\rangle_{\omega}=\frac{M_{\mathbf{q}}}{\omega_{\mathbf{q}}^{2}-\omega^{2}},\label{eq_gf}
\end{equation}
with the spin-wave excitation dispersion relation 
\begin{eqnarray} \label{omega}
&&\omega_{{\bf q}}^{2}  = J_{1}^{2}z_{1}(\gamma_{\mathbf{q}}^{(1)}\hspace{-0.5mm} - \hspace{-0.5mm}1)
\alpha\left(c_{001}\hspace{-0.5mm} - \hspace{-0.5mm}c_{002}\hspace{-0.5mm} -
\hspace{-0.5mm}3\left(c_{101}\hspace{-0.5mm} +
\hspace{-0.5mm}c_{121}\right)\hspace{-0.5mm} + \hspace{-0.5mm}z_{1}c_{001}\gamma_{\mathbf{q}}^{(1)}\right)
\nonumber \\
 && -  (J_{1}^{2}z_{1}(\gamma_{\mathbf{q}}^{(1)}-1)+J_{2}^{2}z_{2}(\gamma_{\mathbf{q}}^{(2)}-1))\left(\lambda c_{001}+\frac{2}{3}S(S+1)\right)\nonumber \\
 && +  J_{2}^{2}z_{2}(\gamma_{\mathbf{q}}^{(2)}-1)\alpha\left(c_{101}-c_{202}-4c_{121}+z_{2}c_{101}\gamma_{\mathbf{q}}^{(2)}\right)\nonumber \\
 && +  J_{1}J_{2}z_{2}(\gamma_{\mathbf{q}}^{(2)}-1)\alpha\left(-4c_{120}-12c_{001}+z_{1}c_{101}\gamma_{\mathbf{q}}^{(1)}\right)\nonumber \\
 && +  J_{1}J_{2}z_{1}(\gamma_{\mathbf{q}}^{(1)}-1)\alpha\left(-3c_{120}-3c_{001}+z_{2}c_{001}\gamma_{\mathbf{q}}^{(2)}\right)\nonumber \\
 && \hspace{-0.0cm}+  J_{1}J_{2}z_{1}z_{2}c_{001}\left(\gamma_{\mathbf{q}}^{(1)}-1\right)\left(\gamma_{\mathbf{q}}^{(2)}-1\right),
\end{eqnarray}
where we have used, on grounds of symmetry, $c_{01-1}\equiv c_{121}$ and $c_{11-1}\equiv c_{221}\equiv c_{120}$.

The correlation functions $c_{\bm{R}}=\frac{1}{N}\sum_{\mathbf{q}}c_{\mathbf{q}}\text{e}^{i\bm{qR}}$
are determined by the spectral theorem,\cite{Tya67} 
\begin{equation}
c_{\mathbf{q}}=\langle S_{\mathbf{q}}^{+}S_{-\mathbf{q}}^{-}\rangle=\frac{M_{\mathbf{q}}}{2\omega_{\mathbf{q}}}[1+2n(\omega_{\mathbf{q}})],\label{eq_C_q}
\end{equation}
where $n(\omega)=(\text{e}^{\omega/T}-1)^{-1}$ is the Bose-Einstein
function. Taking the on-site correlator $c_{\bm{R}=\bm{0}}$ and using
the operator identity \eqref{eq:op_ident}, %
we get the sum rule 
\begin{equation}
\frac{1}{N}\sum_{\mathbf{q}}c_{\mathbf{q}}=\frac{2}{3}S(S+1).\label{eq_sr}
\end{equation}
To calculate the uniform static spin susceptibility
$\chi_{0}=\chi_{\mathbf{q}={\bm 0}}(\omega=0)$,
we use  the relation $\chi_{\mathbf{q}}(\omega)\equiv\chi_{\mathbf{q}}^{zz}(\omega)=\frac{1}{2}\chi_{\mathbf{q}}^{+-}(\omega)$.
In the limit $\omega=0$ and $\mathbf{q}={\bm 0}$ we get from Eq.~(\ref{eq_gf})
\begin{equation} \label{chi_0}
\chi_{0}=\frac{c_{001}J_{1}+c_{101}J_{2}}{\triangle_{\chi_{0}}},
\end{equation}
where 
\begin{align}
& \triangle_{\chi_{0}} =J_{1}^{2}\alpha(9c_{001}-c_{002}-3(c_{101}+c_{121}))-J_{1}J_{2}\alpha(c_{001}-8c_{101}+7c_{120})+J_{2}^{2}\alpha(7c_{101}-4c_{121}-c_{220})\nonumber\\
& -\lambda\left(J_{1}^{2}c_{001}+J_{2}^{2}c_{011}\right)- \frac{2}{3}S(S+1)\left(J_{1}^{2}+J_{2}^{2}\right).
\label{Delta}
\end{align}
The phase with magnetic long-range order at $T \leqslant T_C$ 
is described by the divergence of the static susceptibility 
at the ferromagnetic ordering vector $\bm{Q}_0=\bm{0}$, i.e., by
$\chi_{0}^{-1}=0$.    
In the long-range ordered phase the correlation function $c_{\bm{R}}$
is written as \cite{ShiTak1991,winterfeldt97,RGMlayeredheisenbergarbitraryspin,RGMmoritzcollinearstripe}
\begin{equation}
c_{\bm{R}}=\frac{1}{N}\sum_{\mathbf{q}(\neq\mathbf{0})}c_{\mathbf{q}}\text{e}^{i\bm{qR}}+C,\label{eq_C_R}
\end{equation}
where $c_{\mathbf{q}}$ is given by Eq.~(\ref{eq_C_q}). The condensation
term $C$ determines the magnetization $M$ which is given in
the spin-rotationally invariant form by
\begin{equation}
M^{2}=\frac{3}{2N}\sum_{\bm{R}}c_{\bm{R}}=\frac{3}{2}C.\label{eq:order_parameter}
\end{equation}
According to Eq.~(\ref{chi_0}) the condition for ferromagnetic  long-range
order is
finally given by   $\triangle_{\chi_{0}}=0$, where $\triangle_{\chi_{0}}$
is defined in Eq.~(\ref{Delta}).
The magnetic correlation length above $T_{C}$ is obtained by
expanding $\chi_{\mathbf{q}}$ in the neighborhood of the vector
$\mathbf{Q}_{0}$ \cite{kondoyamaji,winterfeldt97,RGMmoritzcollinearstripe}.
For the ferromagnet we have $\mathbf{Q}_{0}=\bm{0}$, and  we find 
\begin{align}
\xi^{2} &
=\frac{A+B-16\alpha\left(c_{001}^{2}J_{1}^{3}+c_{101}^{2}J_{2}^{3}\right)}{16\left(c_{001}J_{1}+c_{101}J_{2}\right)\triangle_{\chi_{0}}},\label{eq_fm_ksi}\\
A & =-{J_{1}^{2}J_{2}}\alpha\Big(12c_{001}^{2}+c_{001}(31c_{101}-4c_{120})+c_{101}\big(c_{002}+3(c_{101}+c_{121})\big)\Big)\nonumber\\
& -c_{101}\left(\lambda c_{001}+\frac{2}{3}S(S+1)\right),\\
B & =-3J_{1}J_{2}^{2}\alpha\Big(c_{001}(36c_{101}-4c_{121}-c_{220})+c_{101}(16c_{101}+3c_{120})\Big)\nonumber \\
& +c_{001}\left(\lambda c_{011}+\frac{2}{3}S(S+1)\right).
\end{align}
To evaluate the thermodynamic properties, we have to determine the correlation functions
$c_{\bm{R}}$, the vertex parameters $\alpha$ and $\lambda$, and, for $T<
T_C$, the condensation term $C$.
For the correlation functions we use Eq.~(\ref{eq_C_R}) in combination with
Eq.~(\ref{eq_C_q}).
Further we use the sum rule (\ref{eq_sr})  
and the condition for  long-range order, $\triangle_{\chi_{0}}=0$, relevant for $T\le
T_C$. That is sufficient for $S=1/2$, where the vertex parameter $\lambda$ is
not present. 
For $S>1/2$ we need one
more equation. For $T=0$ we use the known 
correlation functions of the
ferromagnetic ground state,    
$c_{\bm{R}}(T=0)=\frac{2}{3}S\delta_{\bm{R},\bm{0}}+\frac{2}{3}S^{2}$, to obtain  $\alpha(T=0)=3/2$ and
$\lambda(T=0)=2-1/S$.
To find the missing equation for $S>1/2$ and $T>0$, we follow  
Refs.~\cite{RGMSgg1anisotropy,heisenbergferro2d,RGMlayeredheisenbergarbitraryspin}
and consider
the ratio 
\begin{equation}
r(T)\equiv
\frac{\lambda(T)-\lambda_{\infty}}
{\alpha(T)-\alpha_{\infty}}=r(0)
\label{eq_r_lambda}
\end{equation}
as independent of temperature.
Here $\lambda_{\infty}$ and $\alpha_{\infty}$ are the vertex parameters for
$T \to \infty$ which can be easily determined as
$\lambda_{\infty}=1-3[4S(S+1)]^{-1}$ and  $\alpha_{\infty}=1$.
To solve the set of RGM equations numerically we use Broyden's method,\cite{NR3}
which yields the
solutions with a   
relative error of about $10^{-8}$ on the average. The momentum integrals are
done by Gaussian integration.

In the low-temperature limit, several analytical expressions can be found
from the RGM equations which are 
identical to those obtained by the linear spin-wave theory (cf., e.g.,
Ref.~\cite{QuantumTheoryoMagn}):
The dispersion relation at zero temperature  and
$|\mathbf{q}|\ll 1$ (long-wavelength
limit) is given by
$\omega_{\mathbf{q}}(T=0)=\rho_{s}(0)\mathbf{q}^{2}$,
where $\rho_{s}(0)=\rho_{s}(T=0)=S|J_{1}+J_{2}|$ is the spin stiffness.
Note that, by contrast to the one- and two-dimensional frustrated $J_1$-$J_2$
Heisenberg
ferromagnet,\cite{RGMchainfrusferro,RGM2Dj1j2frusferro,RGMchainarbitraryspin}
the stiffness remains finite at the classical transition point
$J_2^c=\frac{2}{3}|J_1|$.
The spin stiffness also determines the low-temperature regime of the
magnetization and the specific heat,
\begin{equation}
\frac{M}{S}\sim1-\frac{\zeta(\frac{3}{2})}{2S}\left(\frac{T}{4\pi\rho_{s}(0)}\right)^{\frac{3}{2}},\label{eq:mag_lowT}
\end{equation}
\begin{equation}
C_{V}\sim\frac{15\zeta(\frac{5}{2})}{8}\left(\frac{T}{4\pi\rho_{s}(0)}\right)^{\frac{3}{2}},
\label{eq:cv_lowT}
\end{equation}
where $\zeta(s)$ denotes the Riemann zeta function.

Some remarks are in order here with respect to the comparison between
the RGM and the standard random-phase
approximation (RPA), see, e.g.,
Refs.~\cite{heisenbergferro2d,Tya67,du,froebrich2006,bcc_RPA_2014}.
The spin-wave 
excitation energies calculated within the RGM, see Eq.~(\ref{omega}), exhibit a temperature renormalization 
that is
wavelength dependent and proportional to the correlation functions, so that the
existence of spin-wave excitations does not imply 
a finite magnetization. That is in contrast to
the RPA, where the temperature renormalization of the
spin-wave excitations
is independent of wavelength and proportional to
the magnetization, see, e.g., Refs.~\cite{Tya67,gasser}. 
Correspondingly,
in our theory the stiffness becomes temperature dependent
$\rho_{s}(T)=\sqrt{\alpha(J_{1}+J_{2})\left(c_{001}J_{1}+c_{101}J_{2}\right)}$
due to the temperature dependences of $\alpha$, $c_{001}$, and  $c_{101}$.
Moreover, it is known\cite{heisenbergferro2d,Tya67,gasser}
that the RPA fails in describing magnetic excitations and magnetic short-range order
for $T>T_C$, reflected, e.g., in the
specific heat. On the other hand,  the magnetization for $T\le T_C$ is typically  well
described within RPA.

\subsection{High-temperature series expansion\label{HTE}}

A very general approach to calculate thermodynamic quantities at high and
moderate temperatures is  the high-temperature expansion (HTE)
technique \cite{domb_green,OHZ06,HTE0,HTE,HTE2}.
Here we use a recently developed tool\cite{web_HTE} to calculate the HTE
series of the specific heat and the 
uniform static susceptibility of  model (\ref{eq_ham}) up to order ten.
The region of validity of the HTE series can be extended by  Pad\'e
approximants \cite{domb_green,OHZ06,HTE0,HTE,HTE2,baker61}.
The Pad\'e
approximants $[m,n]$ are ratios of two
polynomials $[m,n]=P_m(x)/R_n(x)$ of degree $m$ and $n$ and they provide an
analytic continuation of a function $g(x)$ given by a power series.
As a rule, Pad\'e approximants with $m \sim n$ provide the best results.

We can use the HTE series for the susceptibility 
$\chi_0=\sum_{n} c_n T^{-n}$
to calculate the Curie
temperature.
Here we use two variants to extract $T_C$ from the HTE series.

(i) We analyze the quotient $q_n=c_n/c_{n-1}$, see, e.g.,
Refs.~\cite{HTE2,wood1955,wood1958,yeomans}. 
For a three-dimensional ferromagnet the critical  behavior of
$\chi_0$ is given by $\chi_0(T) \propto (T-T_C)^{-\gamma}$. 
The expansion of $(T-T_C)^{-\gamma}$ in powers of $1/T$ yields for
the quotient  
$q_n= T_C + (\gamma-1)\frac{T_C}{n}$. For higher orders $n$ 
the HTE series of the model (\ref{eq_ham}) obeys this relation, supposed that a
finite critical temperature exists.
We made a fit of our HTE data
for $q_n$ linearly in  $1/n$ including data points for $n=5,\ldots,10$ to get 
approximate values for $T_C$ and the critical index $\gamma$.
The mean square deviation of the linear fit provides information on the
reliability of this estimate.

(ii) A more sophisticated and more powerful approach to determine $T_C$ and $\gamma$  is based on the so-called differential
approximants (DA) \cite{DA1,DA2,DA3,GUTTMANN2,GUTTMANN3}. Within this
method one considers  differential equations of the form
$
\sum_{\nu=0}^{K}S_{\nu}(\beta)\chi_0^{(\nu)}(\beta)+Y(\beta)=0,
$
where $S_{\nu}(\beta)$ and $Y(\beta)$ are polynomials in $\beta=1/T$
and $\chi_0^{(\nu)}$ denotes the $\nu$-th derivative 
of the susceptibility $\chi_0$. (Note that the
usual Pad\'e approximants follow from $K=0$.) 
The  polynomials $S_{\nu}(\beta)$ and $Y(\beta)$ can be chosen arbitrarily with
the constraint that the total number of free coefficients of all polynomials is equal to
the highest order $n$ of the HTE series (here $n=10$).    
The  coefficients of the polynomials can be determined by inserting the 
derivatives of the HTE series of $\chi_0$ in the equation given above which
yields a simple linear system of equations for the coefficients.   
Assuming critical behavior
for the susceptibility $\chi_0$, we can determine the Curie temperature
$T_C=1/\beta_C$
from $S_{K}(\beta_C)=0$ and the
critical exponent $\gamma$ is given by  $\gamma=K-1-\frac{S_{K-1}(\beta_C)}{S'_{K}(\beta_C)}$.
By varying the degree
of the polynomials $S_{\nu}(\beta)$ and $Y(\beta)$ a set of 21 different 
differential equations was obtained which yields  a corresponding set of
values for $T_C$ and $\gamma$.
However, some of the differential equations lead to unphysical results
(negative or even complex solutions for $T_C$) and have to be discarded.  
The average of the data for  $T_C$ and $\gamma$ obtained from the remaining differential
equations 
yields typically precise
estimates of $T_C$ and $\gamma$, where the mean square deviation provides a
measure of the accuracy (for more details see
Refs.~\cite{DA1,DA2,DA3,GUTTMANN2,GUTTMANN3}).
It is in order to mention here that the HTE naturally works well only if
the Curie temperature is not too small. In particular, for increasing
frustration $J_2$ towards the critical value $J_2^c$ one may expect that the
critical temperature becomes small, and then the determination of $T_C$ and
$\gamma$ from the
HTE series fails.   

We mention that in Ref.~\cite{bcc_oitmaa2004} the HTE series for the staggered
susceptibility was used to determine the N\'{e}el temperature $T_N$ as
a function of $J_2/J_1$ for the model (\ref{eq_ham}) with antiferromagnetic
$J_1$. Moreover, the Curie temperature of the unfrustrated BCC ferromagnet
(i.e.,
$J_2=0$)
for spin quantum numbers $S=1/2,1$ and $3/2$ was calculated from the HTE
series in Ref.~\cite{oitmaa1995_2004}.

\section{Zero-temperature properties\label{sec:GS}}

\begin{figure}
\centering \includegraphics[scale=1]{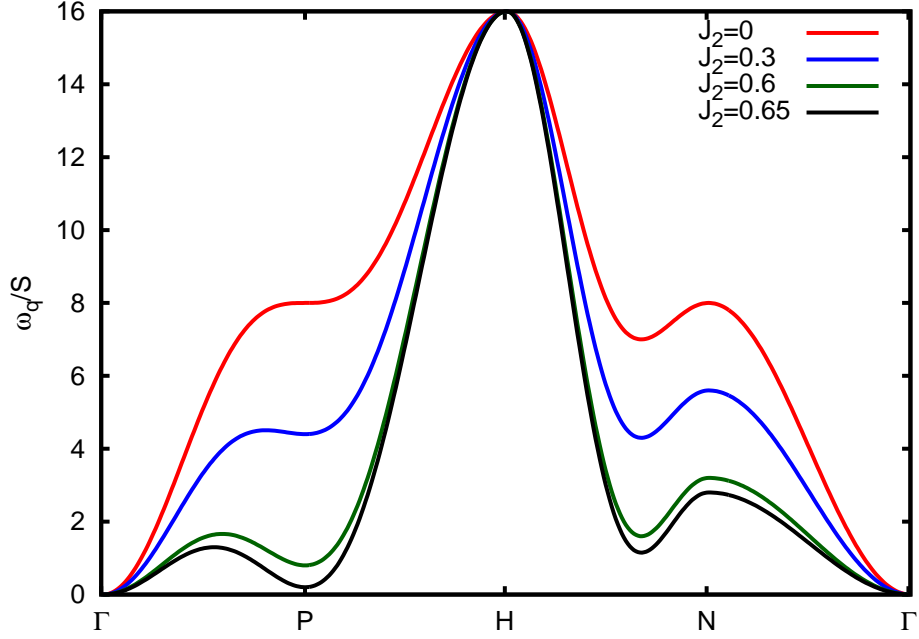} \protect\caption{Dispersion
relation  of the spin-waves at $T=0$ and for different
values of $J_{2}$ and fixed $J_1=-1$. The points in the Brillouin zone are
defined as ${\bf \Gamma}=(0,0,0), {\bf P}=(\pi,\pi,\pi), {\bf H}= (0,0,2\pi),
{\bf N}=(\pi,\pi,0)$. Note that for $T=0$ the dispersion $\omega_{\mathbf{q}}/S$ is independent of
$S$.
}
\label{fig_1} 
\end{figure}

First we use the RGM formalism to study the zero-temperature  properties. 
For $J_2 < J_2^c$ the ground state is the simple ferromagnetic eigenstate
with $\langle \bm{S}_{\bm 0} \bm{S}_{\bf R} \rangle = S^2$ ($\bf R \ne 0$).
The excitation spectrum 
$\omega_{\mathbf{q}}(T=0)$ depends on $J_2$,  see Fig.~\ref{fig_1}.
Since the spin stiffness is given by $\rho_{s}(0)=S|J_{1}+J_{2}|$, see
Sec.~\ref{RGM}, 
the excitation energy in the long-wavelength limit is
reduced by frustration, but the leading $\mathbf{q}^2$-term
in $\omega_{\mathbf{q}}$ remains finite even at $J_2 = J_2^c$.
In the classical model the transition to the collinear antiferromagnetic ground state with
magnetic wave
vector ${\bf Q}=(\pi,\pi,\pi)$ is related by the emergence of a soft mode
at ${\bf q}={\bf Q}$, i.e., $\omega_{\mathbf{Q}}=0$ at
$J_2=J_2^{c,{\rm clas}}$, for ferro- as well as for antiferromagnetic $J_1$.   
We mention here the difference to the $J_1$-$J_2$ ferromagnet in dimension
$D=1$
(Refs.~\cite{RGMchainfrusferro,RGMchainarbitraryspin,sirker2011}) and
$D=2$ (Ref.~\cite{RGM2Dj1j2frusferro}).
In $D=1$ there is no soft mode, the spin stiffness   $\rho_{s}$ vanishes at $J_2 =
J_2^c=|J_1|/4$
(i.e., the leading term in  $\omega_{\mathbf{q}}$ is proportional to
$\mathbf{q}^4$ at $J_2 = J_2^c$), the transition point $J_2^c$ is
independent of $S$, and the 
magnetic wave
vector ${\bf Q}$ changes smoothly from ${\bf Q}={\bf 0}$ to ${\bf Q}\ne {\bf
0}$ when crossing the
transition point
$J_2^c$, see also Refs.~\cite{bader,krivnov2007,LC}.
In $D=2$ the behavior is again different. At the classical transition point
$J_2^{c,{\rm clas}}=|J_1|/2$ all excitation energies
$\omega_{\mathbf{q}}$ with $\mathbf{q}=(q,0)$ and  $\mathbf{q}=(0,q)$, $0\le q \le
\pi$, become
soft (i.e., there is a flat part in $\omega_{\mathbf{q}}$). However, this flat
soft part is
irrelevant for the quantum model, since $J_2^{c,{\rm quant}} < J_2^{c,{\rm clas}}$,
i.e., the transition between the ferromagnetic and the 
antiferromagnetic ground states is shifted to smaller frustration strength 
($J_2^{c,{\rm quant}} \sim 0.4 |J_1|$ for $S=1/2$, see
Refs.~\cite{shannon2006,richter2010}).  

\begin{figure}
\centering \includegraphics[scale=1]{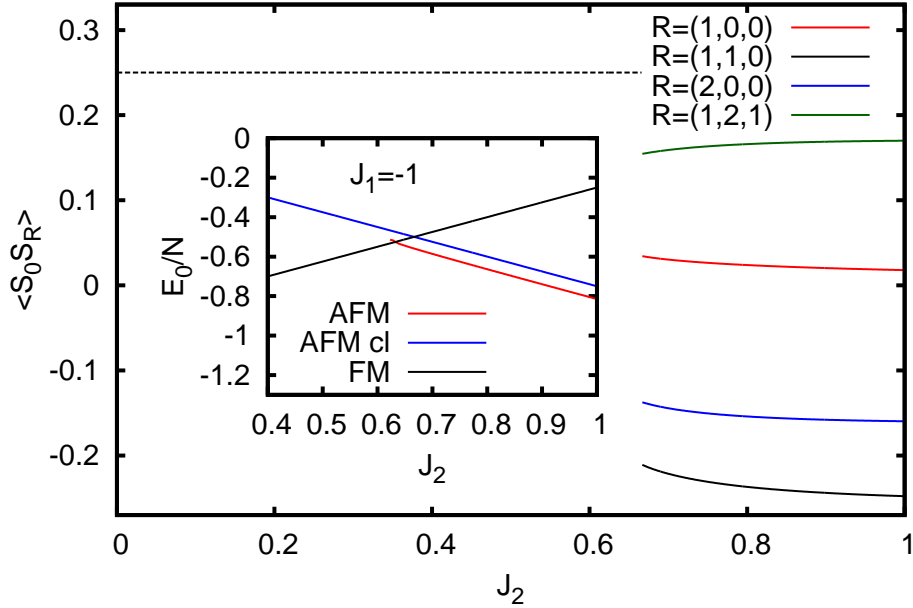}
\protect\caption{
Various ground-state spin-spin correlation functions $\langle {\bf S}_{\bm 0}{\bf S}_{\bf
R}\rangle $  as a function of $J_{2}$ for
the spin-$1/2$ model
in the ferromagnetic ($\mathbf{Q}_{FM}=\mathbf{0}$) and the antiferromagnetic regime
($\mathbf{Q}_{AFM}=\mathbf{P}=\left(\pi,\pi,\pi\right)$). The inset
shows the intersection of the energies per site of the ferromagnetic state and the
collinear antiferromagnetic phase that is used to estimate $J_{2}^{c}$. 
For comparison the GS energy of the classical
model is also shown (labeled by 'AFM cl').
\label{fig:CGSE0}
}

\label{fig_1-1} 
\end{figure}

In view of these results for $D=1$ and $D=2$ the question arises, whether
for $D=3$ the transition point $J_2^{c}$ of the quantum model is different
from that of the classical model. Let us first mention that for the model with
antiferromagnetic NN bond, $J_1>0$, there is a slight shift of $J_2^{c}$ to $J_2^{c}\sim 0.7
J_1 > J_2^{c,{\rm clas}}$
for the $S=1/2$
model \cite{bcc_schmidt2002,bcc_oitmaa2004,bcc_majumdar2009,bcc_RPA_2014}.
To find $J_2^{c}$ for our model with ferromagnetic $J_1$ we extend our RGM approach to the antiferromagnetic region,
i.e., to $
J_2> J_2^{c}$, see Fig.~\ref{fig:CGSE0}. We follow Ref.~\cite{richter2010}
and use the intersection of the
ferromagnetic ground-state energy $E_{FM}/N=-4|J_1| S^2 + 3J_2 S^2$ and the
RGM energy of the collinear antiferromagnetic state to estimate $J_2^{c}$,
see  Fig.~\ref{fig:CGSE0}. For $S=1/2$
we get $J_{2}^{c}\approx 0.63 |J_1|$.
Since the RGM description of the  collinear
antiferromagnetic state with only one vertex parameter $\alpha$ is a poor
approximation,\cite{ShiTak1991,barabanov94,RGMmoritzcollinearstripe}   
a more accurate value for $J_{2}^{c}$ can be obtained using an improved description of the collinear antiferromagnetic ground state
with two vertex parameters $\alpha_1$ and $\alpha_2$, see, e.g.,
Refs.~\cite{winterfeldt97,ihle2001,schmal2006,RGMmoritzcollinearstripe}. The RGM with two
vertex parameters  yields
$J_{2}^{c}\approx 0.68 |J_1|$ for $S=1/2$ and reproduces the classical value
$J_2^{c,{\rm clas}}=2|J_1|/3$ in the large-$S$ limit \cite{unpublished}.
Note that the RPA value for $J_{2}^{c}$, estimated by the
vanishing of $T_C(J_2)$, see the next section, is $J_{2}^{c}=0.6799|J_1|$
\cite{bcc3-1}.

\section{Finite-temperature properties} 
\label{finite_T}

\subsection{Spin-spin correlation functions, magnetization and Curie temperature}
\label{subsec_Tc}

\begin{figure}
\centering 
\includegraphics[scale=1]{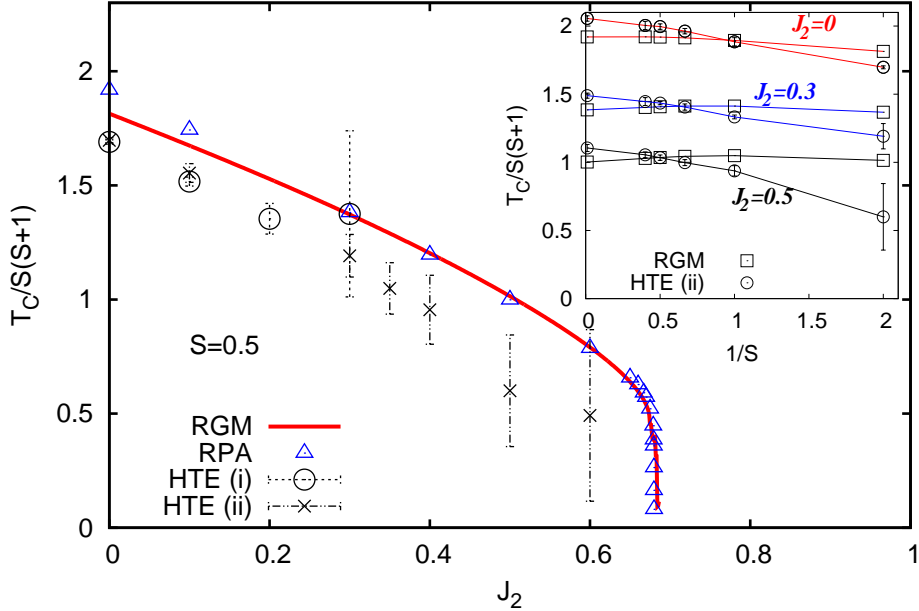} 
\protect\caption{Main panel: Curie temperatures $\frac{T_{C}}{S(S+1)}$ as a function of the
frustration parameter $J_{2}$ and $J_1=-1$ for spin quantum number $S=1/2$
obtained  by the RGM, RPA (data from Ref.~\cite{bcc3-1}), and by the
HTE, variants (i) and (ii).  
Inset: Curie temperatures $\frac{T_{C}}{S(S+1)}$ as a function of $1/S$ for various frustration parameters
$J_{2}$. 
Note that within the RPA approach  $\frac{T_{C}}{S(S+1)}$ is independent of
$S$.
}
\label{fig:Tc} 
\end{figure}

The spin-spin correlation functions $\langle {\bf S}_{\bm 0}{\bf S}_{\bf R}\rangle$,
the magnetization $M$ and the Curie temperature $T_C$ represent the basic
quantities to characterize the ferromagnetic phase.
First we discuss the  Curie temperature $T_C$ as a function of the frustration parameter
$J_2$ as shown in Fig.~\ref{fig:Tc} for $S=1/2$.
(Note that HTE results for the Curie temperature for $S=1/2,1,\ldots,5/2$ and selected values of
$J_2$ are also given in Table~\ref{tab_gamma} in Sec.~\ref{subsec_korrlength}.)     
 We compare the RGM data with the HTE results, variants (i) and (ii), see
Sec.~\ref{HTE}, and the RPA results of Ref.~\cite{bcc3-1}.  Moreover,
for the unfrustrated BCC ferromagnet ($J_2=0$) we can compare with  HTE data of
Ref.~\cite{oitmaa1995_2004} for
$S=1/2$, $1$ and $3/2$.      
 Note that our HTE data for $J_2=0$ within the accuracy of drawing in
 Fig.~\ref{fig:Tc} coincide
with those of Ref.~\cite{oitmaa1995_2004}, where HTE data up to orders
14 ($S=1/2$), 12 ($S=1$), and 9 ($S=3/2$) were used.
For $J_2=0$ and $S=1/2$ the
RGM (RPA) value is about $7\%$ ($13\%$) larger than the HTE value.
The general dependence on $J_2$ is very similar for all methods.
As expected, the error bars of the HTE estimates of  $T_C$  become
larger with increasing $J_2$, and the HTE fails as approaching $J_2^c$,
since the relevant temperatures are too small.
Obviously, the more powerful approach using DA (i.e., HTE variant (ii)) works more reliable until
large values of $J_2$.  
By contrast to the HTE approach,  
the RGM and RPA provide $T_C$ data until the classical transition
point $J_2^{c,{\rm clas}}$. 
As approaching  $J_2^{c,{\rm clas}}$,
the Curie temperature remains quite
large up to about $J_2=0.65$ and finally drops down quickly at  the
transition point.
The fast decay very close to  $J_2^{c,{\rm clas}}$  is well described by a
logarithmic term $1/\textrm{log}(\frac{2}{3}|J_1|-J_{2})$ (cf. also
Refs.~\cite{RGMlayeredheisenbergarbitraryspin} and \cite{yasuda2005}).   
The inset in Fig.~\ref{fig:Tc} demonstrates that the RGM data for the Curie temperature
scaled by $S(S+1)$ are only weakly dependent on $S$. 
The $S$-dependence of the HTE estimate of $T_C/S(S+1)$ is more pronounced,
but it remains also weak (e.g., for $J_2=0.3$ the difference  between the
extreme quantum case $S=1/2$ and the classical limit $S\to \infty$  is about
$20\%$).     
Hence, the curves for
$S=1/2$ shown
in the main panel of  Fig.~\ref{fig:Tc} may be more or less considered as representative
for arbitrary $S$. 
A good description of the RGM data for $T_{C}/S(S+1)$ in the
entire
range $0\le J_2 \le \frac{2}{3}|J_1|$ is provided by the fit
function 
\begin{equation}
\frac{T_{C}}{S(S+1)}=\frac{A(S)}{B(S)-\textrm{log}(\frac{2}{3}|J_1|-J_{2})}.\label{eq:TcJ2BCC-1}
\end{equation}
The fit parameters $A$ and $B$ as a function $S$ are
given  in Table~\ref{tab_3}. Obviously,  the $S$-dependence of $A$ and $B$ is
weak.

\begin{table*}
\protect\caption{Coefficients $A$ and $B$ of the empirical fit function of
the Curie temperature $T_C/S(S+1)$, see  Eq.~(\ref{eq:TcJ2BCC-1}), for various spin quantum numbers $S$.}
\begin{centering}
\begin{tabular}{|c|cccccc|}
\hline 
\phantom{n}\phantom{n}  & \phantom{n} $S=1/2$ \phantom{n}  & \phantom{n} $S=1$ \phantom{n}  & \phantom{n} $S=3/2$ \phantom{n}  & \phantom{n} $S=2$ \phantom{n}  & \phantom{n} $S=5/2$ \phantom{n}  & \phantom{n} $S=100$ \phantom{n} \tabularnewline
\hline 
$A$  & 3.29  & 3.42  & 3.34  & 3.27  & 3.22  & 3.04 \tabularnewline
$B$  & 1.41  & 1.41  & 1.35  & 1.31  & 1.28  & 1.18 \tabularnewline
\hline 
\end{tabular}
\par\end{centering}

\label{tab_3} 
\end{table*}

\begin{figure}
\centering\includegraphics[scale=1]{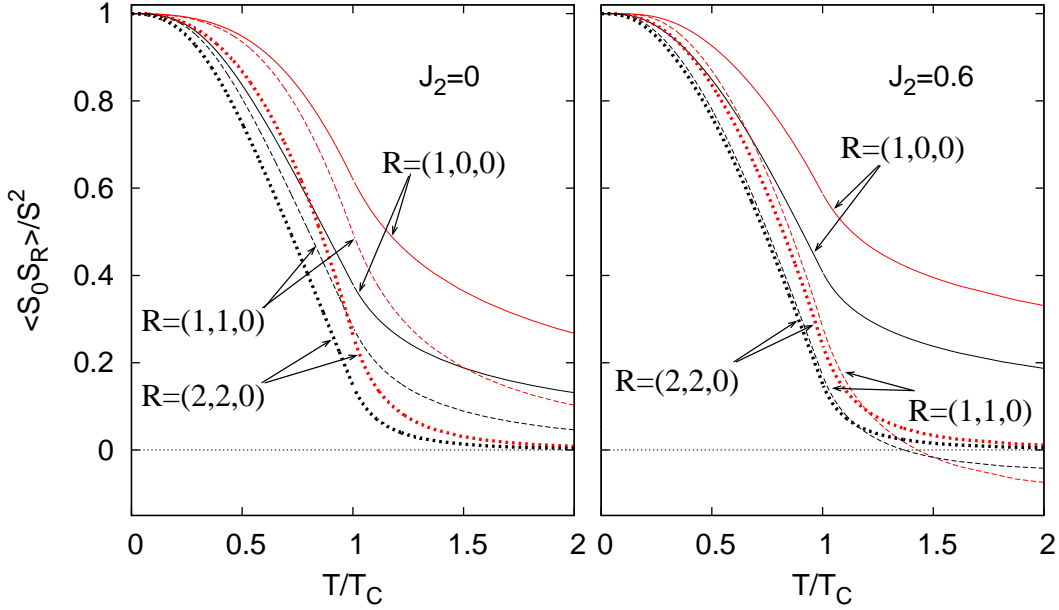}\\
\protect\caption{Spin-spin correlation functions $\langle {\bf S}_{\bm 0} {\bf
S}_{\bf R} \rangle/S^2$ as a function of the normalized temperature $T/T_{C}$
for spin quantum numbers $S=1/2$ (red lines) and $5/2$ (black lines) and for two values of the  frustration
strength
$J_{2}=0$ (left) and $J_{2}=0.6$ (right).
}
\label{fig_sisj} 
\end{figure}

Next we consider the spin-spin correlation functions $\langle {\bf S}_{\bm 0} {\bf
S}_{\bf R} \rangle$. In Fig.~\ref{fig_sisj} we compare the 
temperature dependence of the correlation functions between nearest neighbors
(${\bf R}=(1,0,0)$), next-nearest neighbors
(${\bf R}=(1,1,0)$),
and between spins separated by ${\bf R}=(2,2,0)$  for $J_2=0$ and
$J_2=0.6$.
For  $T<T_C$ the behavior of the correlation functions
 for $J_2=0$ and
$J_2=0.6$ is quite similar, although their decay with increasing $T$ is
faster for larger $S$. Interestingly, above $T_C$ the correlation functions
for particular separations ${\bf R}$ for larger values of frustration  become negative
irrespective of the fact that the ground state is still ferromagnetic.    
These negative correlators are  precursors of the collinear antiferomagnetic
N\'{e}el order present for $J_2 > J_2^c$, since they belong to separations ${\bf R}$, e.g. ${\bf R}=(1,1,0)$, connecting spins on
different sublattices of the large-$J_2$ collinear antiferromagnetic  phase, cf.
also
Fig.~\ref{fig:CGSE0}.     
The general behavior of  $\langle {\bf S}_{\bm 0} {\bf S}_{\bf R} \rangle$ 
for $T \ge T_C$ at large separation $|{\bf R}|$ is
given by  
$ \langle {\bf S}_{\bm 0} {\bf S}_{\bf R} \rangle \backsim
e^{-\frac{|\mathbf{R}|}{\xi(T)}}\left(|\mathbf{R}|^{-D+2-\eta}\right)$,
where $D=3$
and the critical exponent $\eta$ is close to zero,
see, e.g., Refs.~\cite{schneeweis1965,bowers1969,janke1993}. The
long-distance behavior of RGM correlation functions at $T=T_C$ turns out to fit well to
a power law with $\eta \approx 0$ for arbitrary $S$ and $J_2$.

\begin{figure}
\centering\includegraphics[scale=1]{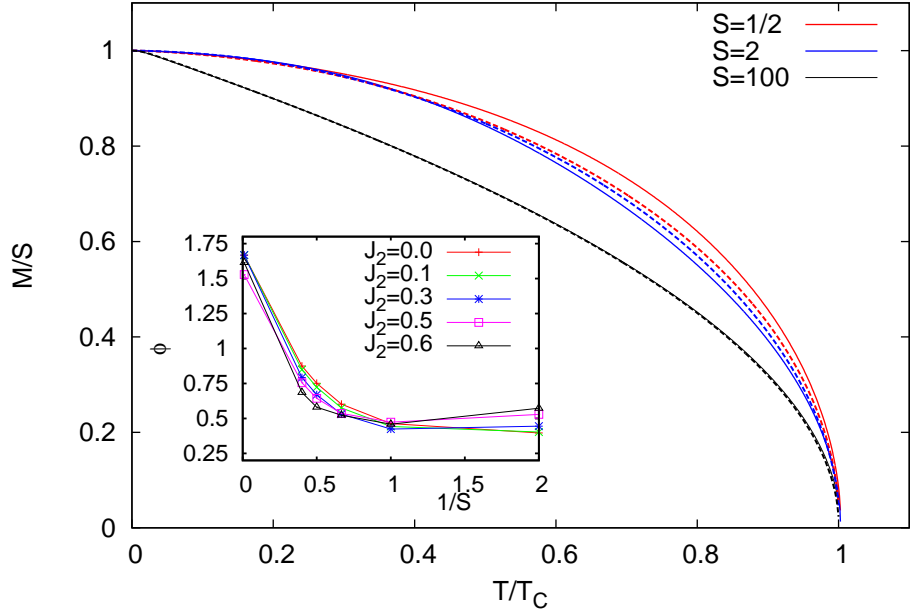}
\protect\caption{Main panel: Magnetization $M/S$ as a function of the normalized temperature $T/T_{C}$
for spin quantum numbers $S=1/2,2$ and $100$ and for two values of the  frustration
strength
$J_{2}=0$ (solid lines) and $J_{2}=0.6$ (dashed lines).
Inset: Shape parameter $\phi$  of the fitting function
(\ref{eq:shape_par_mag}) in dependence on $1/S$
for various values of frustration $J_2$.  
}
\label{M_T} 
\end{figure}

In Fig.~\ref{M_T} we show the temperature dependence of the magnetization 
for three values of $S$ and for two
values of the frustration parameter, $J_2=0$ and
$J_2=0.6$.
The influence of  $J_2$ on the $M/S$
vs. $T/T_C$ curve is rather small, an increase of the spin quantum numbers $S$ leads to
a flattening of the $M/S$
vs. $T/T_C$ curve.   

According to Refs.~\cite{Kuzmin} and \cite{ManRichShapeFM} 
the spontaneous magnetization of a ferromagnet
can be described by the formula 
\begin{equation} 
\frac{M}{S}=\left[1-\phi\left(\frac{T}{T_{C}}\right)^{1.5}-(1-\phi)\left(\frac{T}{T_{C}}\right)^{2.5}\right]^{\beta},\label{eq:shape_par_mag}
\end{equation}
where $\phi$ is the so-called shape parameter and $\beta$ is the
critical index of the order parameter, which is $\beta = 1/2$ within our
RGM approach, cf., e.g.,
Ref.~\cite{RGMlayeredheisenbergarbitraryspin}.  
This formula yields the correct low-temperature behavior and a critical
behavior for $T \to T_C$ with the exponent $\beta$.
Indeed, a corresponding fit of our RGM data for $M$ using the shape
parameter $\phi$ as a single fit parameter yields a very good description of $M(T)$
in the entire temperature region $0\le T \le T_C$.  
We show the shape parameter $\phi$ in the inset of Fig. \ref{M_T}.
Based on the low-$T$ properties known from spin-wave theories and on early
estimates of $T_C$,\cite{bcc4} Kuz'min et al.\cite{Kuzmin,ManRichShapeFM} argued that the dependence of
$\phi$ 
on $S$ and $J_2$ should be weak. Moreover, they found a relation
$0<\phi<5/2$.
Our results for  $\phi$ support these statements.

\begin{figure}[H]
\centering 
\includegraphics[scale=1]{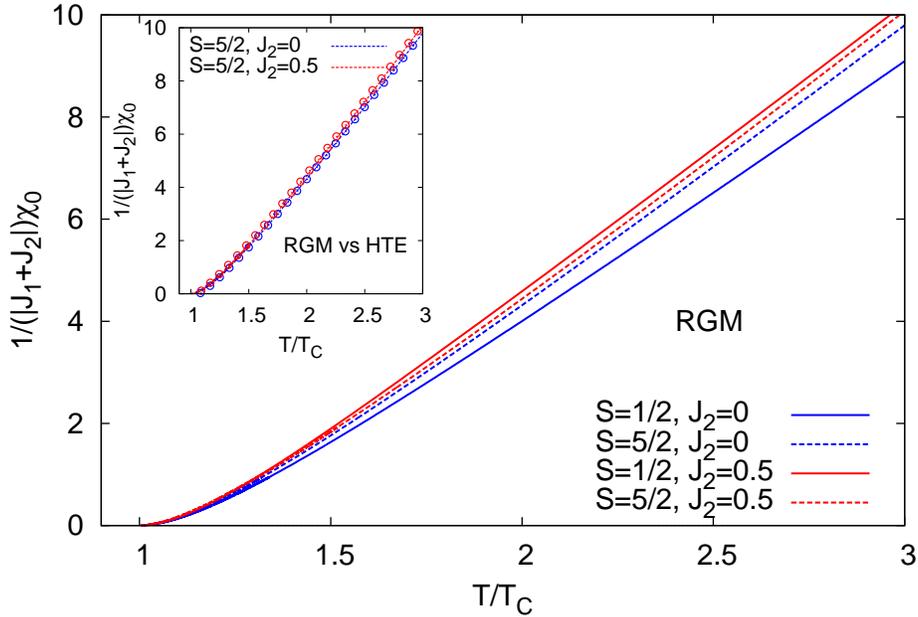}\protect\caption{Inverse scaled susceptibility
$1/(|J_{2}+J_{1}|)\chi$ as a function of the
normalized temperature $T/T_{C}$. Main panel: RGM
results for
$S=1/2$ (solid lines) and $S=5/2$ (dashed lines) for frustration parameters $J_{2}=0$
(blue lines) and
$J_{2}=0.5$ (red lines).
Inset: Comparison of  RGM data (lines) with Pad\'e
approximants $[6,4]$ of the 10th order HTE data (symbols).  
}
\label{fig_4-1} 
\end{figure}

\begin{table*}
\protect\caption{Critical exponent $\gamma$ of the susceptibility and
Curie temperature $T_{C}/S(S+1)$
obtained from HTE (i) and HTE (ii), cf. Sec.~\ref{HTE}, for different spins $S$
and frustration parameters $J_2$ and fixed $J_1=-1$.
}
\selectlanguage{ngerman}%
\noindent \begin{centering}
\begin{tabular}{|c|c|c|c|c|c|c|}
\hline 
\selectlanguage{english}%
$\; \;\; \;\gamma \; \;\;$  \selectlanguage{ngerman}%
 & \multicolumn{1}{c}{$J_{2}=$} & $0$ & \multicolumn{1}{c}{$J_{2}=$} & $0.1$ & \multicolumn{1}{c}{$J_{2}=$} & $0.3$\tabularnewline
\hline 
$S$ & HTE (i) & HTE (ii) & HTE (i) & HTE (ii) & HTE (i) & HTE (ii)\tabularnewline
\hline 
\hline 
\selectlanguage{english}%
$1/2$\selectlanguage{ngerman}%
 & \selectlanguage{english}%
$1.53\pm0.05$\selectlanguage{ngerman}%
 & $1.34\pm0.04$ & \selectlanguage{english}%
$1.51\pm0.10$\selectlanguage{ngerman}%
 & $1.33\pm0.08$ & \selectlanguage{english}%
$0.26\pm2.07$\selectlanguage{ngerman}%
 & $1.38\pm0.40$\tabularnewline
\hline 
$1$ & \selectlanguage{english}%
$1.49\pm0.04$\selectlanguage{ngerman}%
 & $1.44\pm0.20$ & \selectlanguage{english}%
$1.42\pm0.10$\selectlanguage{ngerman}%
 & $1.42\pm0.08$ & \selectlanguage{english}%
$0.46\pm1.30$\selectlanguage{ngerman}%
 & $1.39\pm0.10$\tabularnewline
\hline 
$3/2$ & \selectlanguage{english}%
$1.46\pm0.04$\selectlanguage{ngerman}%
 & $1.40\pm0.10$ & \selectlanguage{english}%
$1.41\pm0.09$\selectlanguage{ngerman}%
 & $1.41\pm0.07$ & \selectlanguage{english}%
$0.57\pm1.18$\selectlanguage{ngerman}%
 & $1.36\pm0.03$\tabularnewline
\hline 
$2$ & \selectlanguage{english}%
$1.45\pm0.04$\selectlanguage{ngerman}%
 & $1.39\pm0.10$ & \selectlanguage{english}%
$1.39\pm0.09$\selectlanguage{ngerman}%
 & $1.43\pm0.10$ & \selectlanguage{english}%
$0.61\pm1.13$\selectlanguage{ngerman}%
 & $1.40\pm0.10$\tabularnewline
\hline 
$5/2$ & \selectlanguage{english}%
$1.44\pm0.04$\selectlanguage{ngerman}%
 & $1.45\pm0.20$ & \selectlanguage{english}%
$1.39\pm0.09$\selectlanguage{ngerman}%
 & \selectlanguage{english}%
$1.40\pm0.07$\selectlanguage{ngerman}%
 & \selectlanguage{english}%
$0.64\pm1.09$\selectlanguage{ngerman}%
 & $1.39\pm0.07$\tabularnewline
\hline 
\end{tabular}
\par\end{centering}

\noindent \begin{centering}
\begin{tabular}{|c|c|c|c|c|c|c|}
\hline 
\selectlanguage{english}%
$\frac{T_{C}}{ S(S+1)}$\selectlanguage{ngerman}%
 & \multicolumn{1}{c}{$J_{2}=$} & $0$ & \multicolumn{1}{c}{$J_{2}=$} & $0.1$ & \multicolumn{1}{c}{$J_{2}=$} & $0.3$\tabularnewline
\hline 
$S$ & HTE (i) & HTE (ii) & HTE (i) & HTE (ii) & HTE (i) & HTE (ii)\tabularnewline
\hline 
\hline 
\selectlanguage{english}%
$1/2$\selectlanguage{ngerman}%
 & \selectlanguage{english}%
$1.67\pm0.01$\selectlanguage{ngerman}%
 & \selectlanguage{english}%
$1.69\pm0.01$\selectlanguage{ngerman}%
 & \selectlanguage{english}%
$1.50\pm0.02$\selectlanguage{ngerman}%
 & \selectlanguage{english}%
$1.55\pm0.04$\selectlanguage{ngerman}%
 & \selectlanguage{english}%
$1.41\pm0.42$\selectlanguage{ngerman}%
 & \selectlanguage{english}%
$1.19\pm0.09$\selectlanguage{ngerman}%
\tabularnewline
\hline 
$1$ & \selectlanguage{english}%
$1.88\pm0.01$\selectlanguage{ngerman}%
 & \selectlanguage{english}%
$1.88\pm0.03$\selectlanguage{ngerman}%
 & \selectlanguage{english}%
$1.71\pm0.03$\selectlanguage{ngerman}%
 & \selectlanguage{english}%
$1.70\pm0.01$\selectlanguage{ngerman}%
 & \selectlanguage{english}%
$1.55\pm0.29$\selectlanguage{ngerman}%
 & \selectlanguage{english}%
$1.33\pm0.01$\selectlanguage{ngerman}%
\tabularnewline
\hline 
$3/2$ & \selectlanguage{english}%
$1.96\pm0.01$\selectlanguage{ngerman}%
 & \selectlanguage{english}%
$1.96\pm0.02$\selectlanguage{ngerman}%
 & \selectlanguage{english}%
$1.79\pm0.03$\selectlanguage{ngerman}%
 & \selectlanguage{english}%
$1.77\pm0.01$\selectlanguage{ngerman}%
 & \selectlanguage{english}%
$1.60\pm0.27$\selectlanguage{ngerman}%
 & \selectlanguage{english}%
$1.40\pm0.02$\selectlanguage{ngerman}%
\tabularnewline
\hline 
$2$ & \selectlanguage{english}%
$1.99\pm0.01$\selectlanguage{ngerman}%
 & \selectlanguage{english}%
$1.99\pm0.02$\selectlanguage{ngerman}%
 & \selectlanguage{english}%
$1.82\pm0.03$\selectlanguage{ngerman}%
 & \selectlanguage{english}%
$1.80\pm0.02$\selectlanguage{ngerman}%
 & \selectlanguage{english}%
$1.63\pm0.26$\selectlanguage{ngerman}%
 & \selectlanguage{english}%
$1.43\pm0.01$\selectlanguage{ngerman}%
\tabularnewline
\hline 
$5/2$ & \selectlanguage{english}%
$2.01\pm0.01$\selectlanguage{ngerman}%
 & \selectlanguage{english}%
$2.00\pm0.03$\selectlanguage{ngerman}%
 & \selectlanguage{english}%
$1.84\pm0.03$\selectlanguage{ngerman}%
 & \selectlanguage{english}%
$1.82\pm0.01$\selectlanguage{ngerman}%
 & \selectlanguage{english}%
$1.64\pm0.26$\selectlanguage{ngerman}%
 & \selectlanguage{english}%
$1.44\pm0.03$\selectlanguage{ngerman}%
\tabularnewline
\hline 
\end{tabular}
\par\end{centering}

\selectlanguage{english}%
\label{tab_gamma} 
\end{table*}

\subsection{Susceptibility and correlation length}
\label{subsec_korrlength}

Next we discuss the uniform  susceptibility  $\chi_0$ and the correlation
length $\xi$.   In addition to the spin-spin correlation functions, see
Sec.~\ref{subsec_Tc}, both quantities characterize the paramagnetic phase above
$T_C$. 
We show the inverse susceptibility in Fig.~\ref{fig_4-1}. 
The susceptibility multiplied by $(|J_{2}+J_{1}|)$ 
shows an almost universal behavior as a function of the normalized temperature
$T/T_{C}$, i.e., the dependence on $S$ and $J_2$ is small.
The RGM results agree well with the HTE data (see the inset in
Fig.~\ref{fig_4-1}). 
While the RGM critical exponents deviate from the correct values, the HTE
approach provides more accurate exponents. We present HTE data for $\gamma$ in
Table~\ref{tab_gamma}. As expected, $\gamma$ is almost independent of $S$ and
$J_2$ and  it is close to the correct value $\gamma
\sim 1.39$ of the three-dimensional Heisenberg ferromagnet, see, e.g.,
Ref.~\cite{janke1993}.
As already noticed in Sec.~\ref{subsec_Tc}, the  HTE approach (ii) to determine $T_C$ and $\gamma$  using 
differential approximants is more accurate and works well until large
values of $J_2$. Thus, the analysis of the quotient of the HTE series for
$\chi_0$, i.e. HTE (i), starts to fail already at $J_2 \sim 0.3$, see the
large least-square deviations in Table~\ref{tab_gamma}.

\begin{figure}[H]
\centering 
\includegraphics[scale=1]{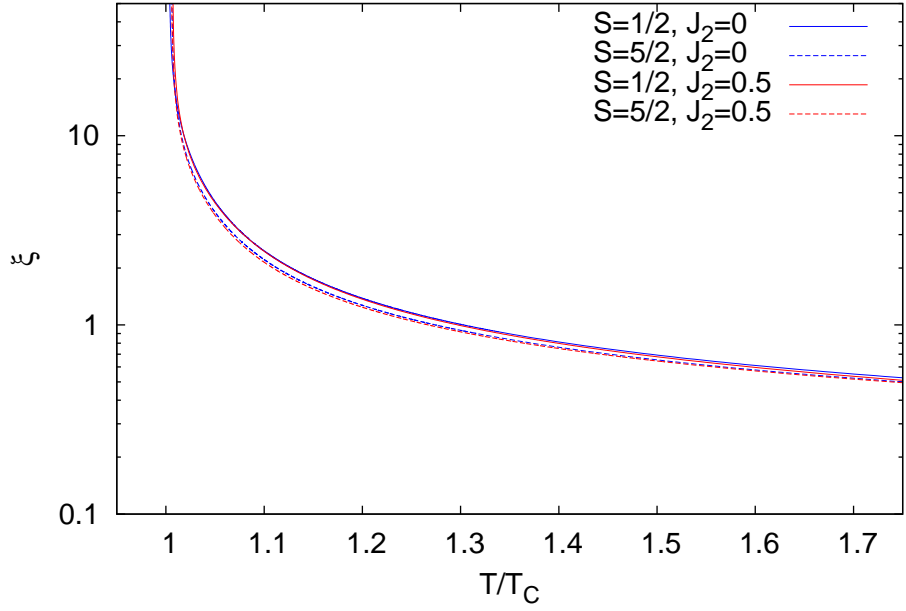}\protect\caption{Correlation length
$\xi$ as a function of the
normalized temperature $T/T_{C}$ for 
$S=1/2$ (solid lines) and $S=5/2$ (dashed lines) for frustration parameters
$J_{2}=0$
(blue lines) and
$J_{2}=0.5$ (red lines).
}
\label{cor_length} 
\end{figure}

The RGM results for the correlation length are shown in
Fig.~\ref{cor_length}. 
Again the influence of the spin quantum number $S$ and the frustration $J_2$
on the $\xi(T/T_C)$-curve 
is weak. The decay of $\xi$ with temperature  is fast, already at $T \sim
1.3 T_C$ the correlation length is of the order of the lattice constant. 

\begin{figure}[H]
\centering \includegraphics[scale=1]{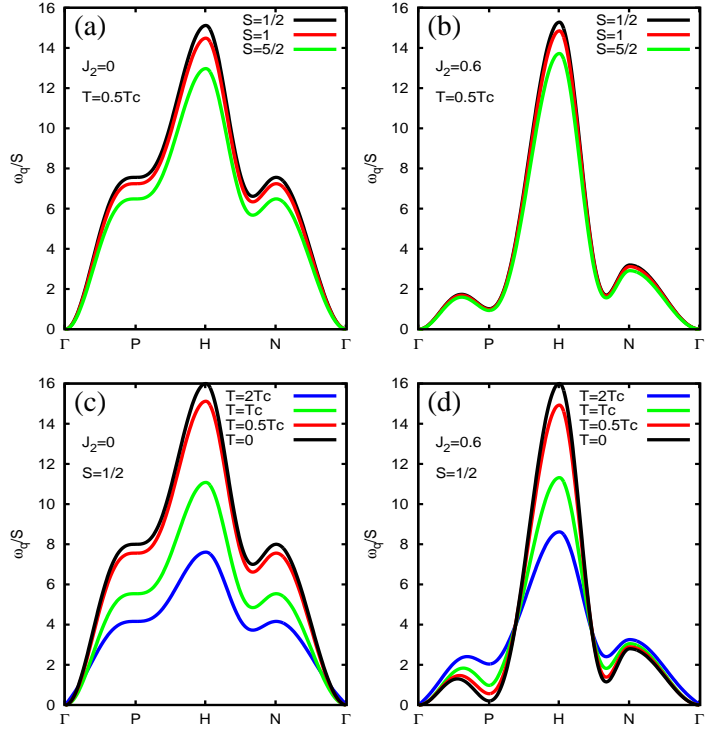}
\protect\caption{Temperature dependence of the dispersion
relation $\omega_{\mathbf{q}}/S$ of the spin excitations  for different
values of $J_{2}$ and fixed $J_1=-1$. The points in the Brillouin zone are
defined as ${\bf \Gamma}=(0,0,0), {\bf P}=(\pi,\pi,\pi), {\bf H}= (0,0,2\pi),
{\bf N}=(\pi,\pi,0)$.}
\label{O_q_T} 
\end{figure}

\subsection{Excitation spectrum and specific heat}
\begin{figure}[H]
\centering\includegraphics[scale=1]{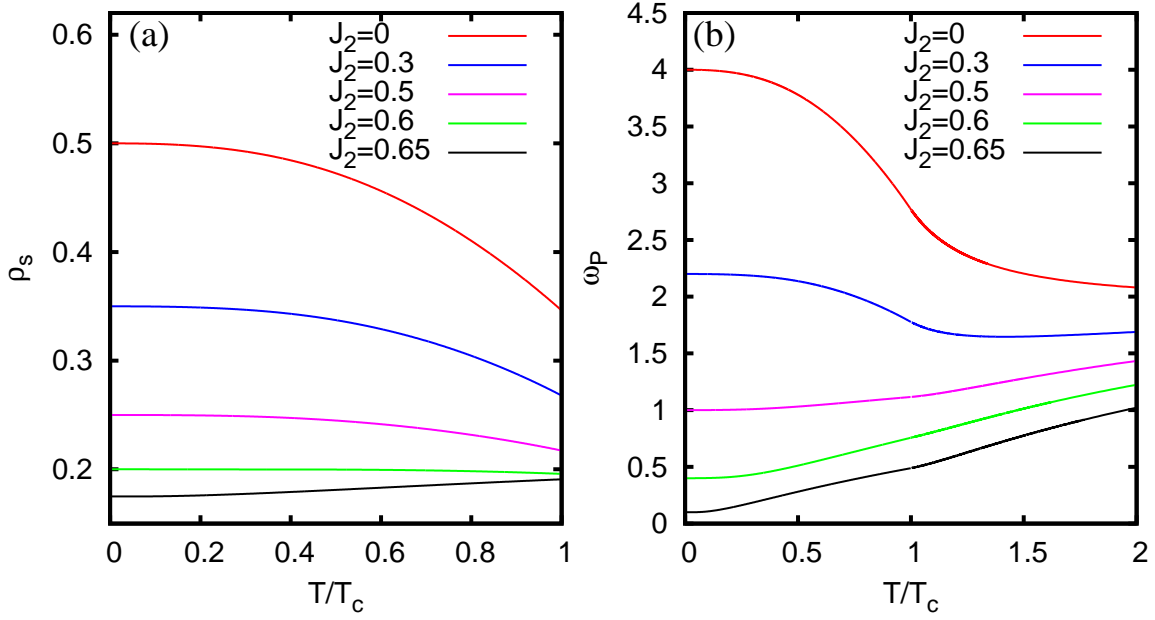}
\protect\caption{Temperature dependence of the spin stiffness $\rho_{s}$
(left) and
excitation energy $\omega_P$ at the soft-mode ${\bf q}$-vector ${\bf P}=(\pi,\pi,\pi)$
(right) for
spin $S=1/2$.
}
\label{fig9} 
\end{figure}

Let us now  consider energetic quantities. By contrast to the RPA, the RGM provides a
reasonable    
description of the excitation spectrum $\omega_{\mathbf{q}}$  for arbitrary
temperatures. 
We show the dispersion
relation  for various temperatures, spin
quantum numbers and frustration parameters in Fig.~\ref{O_q_T}.
First we notice that for finite temperatures   $\omega_{\mathbf{q}}/S$ becomes
dependent
on $S$, see Figs.~\ref{O_q_T}a and b.  It is
also evident that for a wide range of temperatures up to about $50\%$
of the Curie temperature the changes in   $\omega_{\mathbf{q}}$  are quite small, see
also Fig.~\ref{fig9}.
The main aspects  of the influence of the temperature on $\omega_{\mathbf{q}}$
are the following:
For small values of $J_2 \lesssim 0.3$ the excitation energies for all
${\bf q}$-vectors are
reduced by increasing $T$. Around  the maximum at  ${\bf H}= (0,0,2\pi)$ this
reduction is present for all
values of $J_2$. However, the temperature dependence of $\omega_{\mathbf{q}}$  around the
dip/minimum at the soft-mode wave vector    
${\mathbf{q}}= {\bf P}=(\pi,\pi,\pi)$
as well as for ${\mathbf{q}} \to \bm{0}$  
differs for smaller $J_2$ from that 
for larger $J_2$. 
(Note that the second  dip/minimum between the H- and N-points
at about ${\mathbf{q}}=0.66(\pi,\pi,\pi)$  is related to that at
${\mathbf{q}}= {\bf P}$, i.e. it appears only due to the choice of path ${\bf
H} \to {\bf N}$ in the
Brillouin zone.)   
At ${\mathbf{q}}= {\bf P}$
 the excitation energy $\omega_{\bf P}$ 
decreases with growing $T$ for smaller $J_2$,  but it increases for larger $J_2$,
see Fig.~\ref{fig9}b.
The influence of $T$ on  $\omega_{\mathbf{q}}$ near the ${\bf \Gamma}$-point,
i.e., at small
$|\mathbf{q}|$, is more subtle:
For $T<T_C$  the temperature dependence of the excitation energies 
is given by the temperature dependence of the spin stiffness $\rho_s$
describing the quadratic term in  $\omega_{\mathbf{q}}$ for
small $|\mathbf{q}|$, while  for $T>T_C$ a linear term in $\omega_{\mathbf{q}}$
determines the behavior  near the ${\bf \Gamma}$-point. 
We show $\rho_{s}(T)$ for $S=1/2$ for various values of $J_2$ in
Fig.~\ref{fig9}a. As expected for
ferromagnets,\cite{Lovesey1977,Sun2006,Katanin2} 
the stiffness becomes smaller with increasing of $T$ for $J_2 \lesssim 0.6$.
Interestingly, for $J_2 \gtrsim 0.6$
$\rho_{s}(T)$ is growing with $T$.
This unusual behavior of $\rho_{s}(T)$ near the zero-temperature transition
point to an antiferromagnetic state has recently been discussed in
Ref.~\cite{Katanin2} using non-linear and self-consistent spin-wave
theories. Note, however, that  in Ref.~\cite{Katanin2}
the frustrated FCC ferromagnet is considered, where  $\rho_{s}(T=0)=0$ at 
the transition point to the  antiferromagnetic ground state, whereas for our
BCC model
$\rho_{s}(T=0)>0$ at $J_2^c$.
 For small and
moderate frustration, the spin-wave approach of Ref.~\cite{Katanin2} yields a
 decrease of
$\rho_s$ with $T$ according to  
$\rho_s(T)=\rho_s(0)-BT^{\mu}$, $\mu=5/2$ and $B>0$.
As approaching the zero-temperature transition point
to the antiferromagnetic
ground state, the exponent $\mu$ changes to
$\mu=5/4$ and the prefactor $B$ becomes negative, i.e. the stiffness can
grow with increasing $T$.
Our RGM data for $\rho_s(T)$ confirm these predictions:
We find that $\rho_s(T)$ is well described by the  power law given above with
$B>0$ and  $\mu \approx 2.5$ for $J_2 \lesssim 0.5$, whereas for $J_2=0.6$  
we have $B \approx -0.05$ and  $\mu \approx 1.5$.

\begin{figure}[H]
\centering\includegraphics[scale=1]{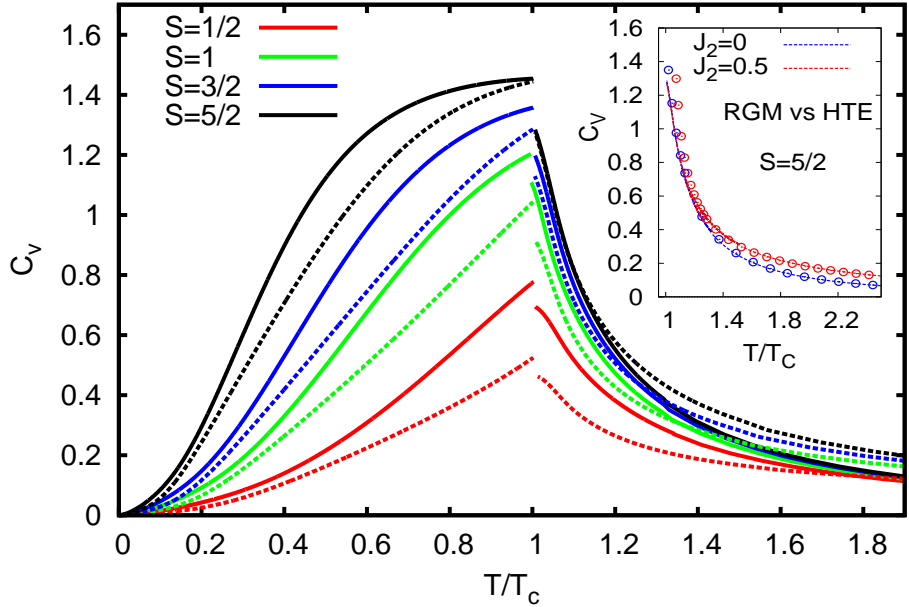} 
\protect\caption{Specific heat  $C_V$ as a function of the normalized temperature
$T/T_{C}$.
Main panel: RGM data for $S=1/2,1,3/2$ and $5/2$  for frustration parameters
$J_{2}=0$ (solid lines) and $J_{2}=0.5$ (dashed lines). 
Inset: Comparison of  RGM data (dotted lines) with Pad\'e
approximants $[6,4]$ of the 10th order HTE data (symbols).  
}
\label{C_V} 
\end{figure}

The specific heat $C_V$ is shown in Fig.~\ref{C_V}.
By contrast to the susceptibility and  the correlation length the influence of
the spin $S$ and the frustration $J_2$ is noticeable.
The $C_V(T)$ curves  show the characteristic cusplike 
behavior at the transition temperature
around $T_{C}$ for small spin quantum numbers $S$ indicating the second-order phase transition. 
With increasing $S$ in the long-range ordered phase 
the slope of the $C_V$ curves near $T_C$ decreases, and the cusplike shape develops
to a kinklike one. This behavior for larger $S$ may be considered as a
shortcoming of the RGM, cf.
Ref.~\cite{RGMlayeredheisenbergarbitraryspin}.
For $T>T_C$ we can  compare the RGM data with HTE results, see the inset in
Fig.~\ref{C_V}. As for the susceptibility, the agreement is very good.
The low-temperature behavior is given by Eq.~(\ref{eq:cv_lowT}), i.e., 
$C_V(T)$ increases with increasing $J_2$. Note, however, that the specific
heat as a function of the normalized temperature
$T/T_{C}$ shows the opposite trend, see Fig.~\ref{C_V}.

\section{Summary}
\label{sec:sum} 
In this paper 
we use the rotation-invariant Green's function
method (RGM) to calculate thermodynamic quantities, such as the Curie temperature
$T_C$, the spontaneous magnetization $M$, the spin-spin correlation
functions
$\langle {\bf S}_{\bf 0} {\bf S}_{\bf R} \rangle$,  the uniform susceptibility
$\chi_0$, the correlation length
$\xi$, the specific heat $C_{V}$  and the spin stiffness  $\rho_s$  of a 
frustrated
spin-$S$ Heisenberg magnet on the BCC
lattice with ferromagnetic NN exchange $J_1$ and antiferromagnetic NNN exchange
$J_2$.
We focus on the ferromagnetic regime, i.e., $J_2 \le 2|J_1|/3$, such that the
ground state is ferromagnetic.
For $T>T_C$ our Green's function approach is complemented  
by 10th order high-temperature expansion for the susceptibility
and the specific heat.  

The RGM  
goes one step beyond the random-phase
approximation (RPA).  As a result, several shortcomings of the RPA, see,
e.g.,
Refs.~\cite{du,Tya67,gasser,froebrich2006,bcc_RPA_2014}, such as  
the artificial equality of the  critical temperatures $T_N= T_C$ for ferro- and
antiferromagnets or the failure in describing the paramagnetic
phase  at $T>T_C$, can be overcome. 
For the Curie temperature and the spontaneous magnetization $M$ 
we derive simple fit formulas describing $T_C$ as a
function of $S$ and $J_2$ and the temperature dependence of $M(S,J_2)$.  
As approaching the ground-state transition point to the antiferromagnetic
phase at $J_2^c \approx 2|J_1|/3$, the thermodynamic properties deviate from
the ordinary ferromagnetic behavior. Thus, the spin-spin correlators 
may become negative at  $T>T_C$ indicating the collinear antiferromagnetic
N\'{e}el order present for $J_2 > J_2^c$, and the temperature profile of
the spin stiffness $\rho_s$ for  $T<T_C$ exhibits an increase with $T$
instead of the ordinary decrease.

The present investigations are focussed on theoretical aspects, in
particular, with respect to previous discussions of one- and two-dimensional
frustrated ferromagnets. There might be some relevance for ferromagnetic
compounds \cite{exp1,exp2,exp3} especially near a quantum phase transition,
e.g., driven by
doping.


\begin{thebibliography}{10}
\bibitem{bader} H.P. Bader and R. Schilling,
Phys. Rev. B {\bf 19}, 3556 (1979).
\bibitem{hamada} T. Hamada, J. Kane, S. Nakagawa, and Y. Natsume, J. Phys. Soc.
Jpn. {\bf 57}, 1891 (1988); {\bf 58}, 3869 (1989).   
\bibitem{shannon2006} N. Shannon,T. Momoi, and P. Sindzingre,
Phys. Rev. Lett. {\bf 96}, 027213 (2006).
\bibitem{krivnov2007}
D. V. Dmitriev, V. Ya. Krivnov, and J. Richter,
Phys. Rev. B \textbf{75}, 014424 (2007).
 
\bibitem{richter2010}
J. Richter, R. Darradi, J.~Schulenburg, D.J.J. Farnell, and H.
       Rosner,
        Phys. Rev. B {\bf 81}, 174429 (2010).

\bibitem{momoi2011} R. Shindou, S. Yunoki, and T. Momoi,
Phys. Rev. B {\bf 84}, 134414 (2011).

\bibitem{LC} V. Dmitriev and V. Ya. Krivnov, Phys. Rev. B {\bf 73}, 024402
(2006),
 A. Nersesyan, A. O. Gogolin, and F. H. L. Essler, 
Phys. Rev. Lett. {\bf 81}, 910 (1998);
C. Cabra, A. Honecker, and P. Pujol, Eur. Phys. J. B {\bf 13}, 55
(2000).


\bibitem{selke77}
W. Selke, Z. Phys. B {\bf 27}, 81 (1977).


\bibitem{tmrg} H. T. Lu, Y. J. Wang, Shaojin Qin, and T. Xiang,
Phys. Rev. B \textbf{74}, 134425 (2006); J.~Sirker, {Phys. Rev. B} {\bf81},
014419 (2010).


\bibitem{RGMchainfrusferro}
M. H\"{a}rtel, J. Richter, D. Ihle, and  S.-L.
Drechsler,
       Phys. Rev. B {\bf 78}, 174412 (2008).


\bibitem{RGM2Dj1j2frusferro}M. H\"artel, J. Richter, D. Ihle, and S.-L.
Drechsler,
 Phys. Rev. B \textbf{81}, 174421 (2010).

\bibitem{dmitriev_clas} D.V. Dmitriev and V.Y. Krivnov,
Phys. Rev. B {\bf 82}, 054407 (2010); Eur. Phys. J. B {\bf 82}, 123 (2011).




\bibitem{RGMchainarbitraryspin}M. H\"artel, J. Richter, D. Ihle, J.
Schnack, and S.-L. Drechsler,
Phys. Rev. B \textbf{84}, 104411 (2011).




\bibitem{sirker2011}  J. Sirker, V. Y. Krivnov, D. V. Dmitriev, A. Herzog, O. Janson, S.
Nishimoto, S.-L. Drechsler,
       and J. Richter,
       Phys. Rev. B {\bf 84}, 144403 (2011).



\bibitem{bcc_schmidt2002}
R. Schmidt, J.~Schulenburg, J. Richter, and D.D. Betts,
    Phys. Rev. B {\bf 66}, 224406 (2002). 

\bibitem{bcc_oitmaa2004} J. Oitmaa and Weihong Zheng,
Phys Rev. B {\bf 69}, 064416 (2004).

\bibitem{bcc_majumdar2009} K. Majumdar and T. Datta, J. Phys.: Condens. Matter {\bf 21}, 406004
(2009).

 
\bibitem{bcc_momoi2013} H.T. Ueda and  T. Momoi,
Phys. Rev. B \textbf{87}, 144417 (2013).

\bibitem{kondoyamaji}J. Kondo and K. Yamaji, 
Prog. Theor. Phys. \textbf{47}, 807 (1972).

\bibitem{RhoScal94} E. Rhodes and S. Scales, 
Phys. Rev. B {\bf  8}, 1994 (1973).

\bibitem{ShiTak1991} H.~Shimahara and S.~Takada, 
J.~Phys.~Soc.~Jpn.~\textbf{60}, 2394 (1991).



\bibitem{SSI94}F. Suzuki, N. Shibata, and C. Ishii,
J. Phys. Soc. Jpn. \textbf{63}, 1539 (1994).

\bibitem{barabanov94}   A.F. Barabanov and V.M. Berezovskii, J. Phys. Soc.
Jpn. {\bf 63}, 3974 (1994); 
Phys. Lett. A {\bf 186}, 175 (1994);  Zh. Eksp. Teor. Fiz. {\bf
106}, 1156 (1994) [JETP {\bf 79}, 627 (1994)].


\bibitem{winterfeldt97} S. Winterfeldt and D. Ihle, Phys. Rev. B
\textbf{56}, 5535 (1997).

\bibitem{SIH00} L. Siurakshina, D. Ihle, and R. Hayn,
Phys. Rev. B \textbf{61}, 14601 (2000).


\bibitem{yu_feng2000} W. Yu and S. Feng, Eur. Phys. J. B {\bf{13}}, 265 (2000); 

\bibitem{ihle2001} L. Siurakshina, D. Ihle, and R. Hayn, Phys. Rev. B
\textbf{64}, 104406 (2001).


\bibitem{canals2002}
B.H. Bernhard, B. Canals, and C. Lacroix, Phys. Rev. B {\bf{66}},
 104424  (2002).
\bibitem{prb2004} D. Schmalfu{\ss}, J. Richter, and D. Ihle, Phys. Rev. B \textbf{70}, 184412 
(2004).

\bibitem{RGMferromagneticfield}I. Juhasz Junger, D. Ihle, J. Richter,
and A. Kl\"umper,
Phys. Rev. B \textbf{70}, 104419 (2004).

\bibitem{RGMSgg1anisotropy}I. Juhasz Junger, D. Ihle, and J. Richter,
Phys. Rev. B \textbf{72}, 064454 (2005).

\bibitem{RGMquasi2Dvskagomelattice}D. Schmalfu{\ss}, J. Richter, and D. Ihle,
Phys. Rev. B \textbf{72}, 224405 (2005).

\bibitem{schmal2006}
       D. Schmalfu{\ss}, R. Darradi, J. Richter, J.~Schulenburg,  and D.~Ihle, 
       Phys. Rev. Lett. {\bf 97}, 157201 (2006). 


\bibitem{antsyg} T.N.~Antsygina,
M.I.~Poltavskaya, I.I.~Poltavsky, and K.A.~Chishko, 
Phys. Rev. B {\bf 77},  024407 (2008).

\bibitem{heisenbergferro2d}I. Juhasz Junger, D. Ihle, L. Bogacz, and
W.Janke,
Phys. Rev. B \textbf{77}, 174411 (2008).

\bibitem{RGMlayeredheisenbergarbitraryspin}I. Juhasz Junger, D. Ihle,
and  J. Richter,
Phys. Rev. B \textbf{80}, 064425 (2009).


\bibitem{RGMchainfrusferromagnet}M. H\"artel, J. Richter, and D. Ihle,
Phys. Rev. B \textbf{83}, 214412 (2011).

\bibitem{antsyg2012}
 I.I.~Poltavsky, T.N.~Antsygina,
M.I.~Poltavskaya, and K.A.~Chishko, 
Physica B: Cond. Mat. {\bf 407},  3925 (2012).



\bibitem{RGMmoritzcollinearstripe} M. H\"artel, J. Richter, O. G\"otze,
D. Ihle, and S.-L. Drechsler,
Phys. Rev. B \textbf{87}, 054412 (2013).



\bibitem{domb_green} G.S.~Rushbrooke, G.A. Baker, and P.J. Wood, in
{\em Phase Transitions and Critical Phenomena}, Vol. 3, p. 245; eds.
C.~Domb and M.S.~Green, Academic Press, London, 1974.


\bibitem{OHZ06}
J.~Oitmaa, C.J.~Hamer, and W.H.~Zheng,
\newblock {\em {Series Expansion Methods}},
Cambridge University Press, Cambridge, 2006.


\bibitem{HTE0} H.-J. Schmidt, J. Schnack, M. Luban,
Phys. Rev. B \textbf{64}, 224415 (2001).

\bibitem{HTE}H.-J. Schmidt, A. Lohmann, and J. Richter,
Phys. Rev. B \textbf{84}, 104443 (2011).

\bibitem{HTE2} A. Lohmann, H.-J. Schmidt, and J. Richter,
Phys. Rev. B \textbf{89}, 014415 (2014).



\bibitem{Tya67} S.~V.~Tyablikov, \textit{Methods in the Quantum
Theory of Magnetism} (Plenum, New York, 1967).


\bibitem{gasser}W. Gasser, E.Heiner, and K. Elk, \textit{Greensche Funktionen
in Festk\"orper- und Vielteilchenphysik}. WILEY-VCH,  2001.


\bibitem{QuantumTheoryoMagn} W. Nolting and A. Ramakanth, \emph{Quantum
Theory of Magnetism} (Springer Berlin, 2010).

\bibitem{NR3} W.H.~Press, S.A.~Teukolsky, W.T.~Vetterling, B.P.~Flannery,
{\it Numerical Recipes in C++, The Art of Scientific Computing}, Cambridge
University Press, Cambridge,
2007.


\bibitem{bcc_RPA_2014} 
M.R. Pantic, D.V. Kapor, S.M. Radosevic, and P.Mali, 
Solid State Comm. {\bf 182}, 55 (2014).

\bibitem{du} A.~Du  and G.Z.~Wei, J. Magn. Magn. Mater. {\bf 137}, 343
(1994). 


\bibitem{froebrich2006}  P. Froebrich and P.J. Kuntz,
Physics Reports {\bf 432}, 223 (2006).





\bibitem{web_HTE} see {\tt http://www.uni-magdeburg.de/jschulen/HTE10/} 
and Ref.~\cite{HTE2}.


\bibitem{baker61}
G.A. Baker, Phys. Rev. {\bf 124}, 768 (1961).


\bibitem{wood1955} G.S. Rushbrooke and P.J. Wood,
Proc. Phys. Soc. A {\bf 68}, 1161 (1955).

\bibitem{wood1958}  G.S. Rushbrooke and P.J. Wood,
Molecular Physics {\bf 1}, 257 (1958).



\bibitem{yeomans}
J. M. Yeomans
{\it
Statistical Mechanics of Phase Transitions},
Oxford University Press, Oxford, 1962.


\bibitem{GUTTMANN2} A.J. Guttmann,
\textit{Asymptotic Analysis of
Power Series Expansions}, Academic Press, 1989.

\bibitem{GUTTMANN3} A.J. Guttmann and G.S. Joyce,
J. Phys. A: Gen. Phys. \textbf{5}, L81 (1972).


\bibitem{DA1}   D.L. Hunter, G.A. Baker, 
Phys. Rev. B \textbf{19}, 3808 (1979).

\bibitem{DA2}Michael E. Fisher and H. Au-Yang,
J. Phys. A: Math. Gen. \textbf{12}, 1677 (1979).

\bibitem{DA3}M. Roger. 
Phys. Rev. B \textbf{58}, 11115 (1998).

\bibitem{oitmaa1995_2004}  J. Oitmaa and E. Bornilla,
Phys. Rev. B {\bf 53}, 14228 (1995);
    J. Oitmaa and Weihong Zheng, J. Phys.: Condens. Matter {\bf 16}, 8653
 (2004).




\bibitem{unpublished} P. M\"uller, J. Richter, and D. Ihle, in preparation.

\bibitem{bcc3-1} R. A. Tahir-Kheli and H. S. Jarrett,
Phys. Rev. \textbf{135}, A1096 (1964).


\bibitem{yasuda2005} C. Yasuda, S. Todo, K. Hukushima, F. Alet, M. Keller, M.
Troyer, and H. Takayama, Phys. Rev. Lett. {\bf 94}, 217201 (2005). 


\bibitem{schneeweis1965}
C. Schneiweiss, Z. f. Phys. {\bf 182}, 466 (1965).

\bibitem{bowers1969}
R. G. Bowers and M. E. Woolf,
Phys. Rev. {\bf 177}, 917 (1969). 

\bibitem{janke1993} C. Holm and W. Janke, 
Phys. Rev. B {\bf 48}, 936 (1993).

\bibitem{Kuzmin} M. D. Kuz'min, Phys. Rev. Lett. {\bf 94}, 107204 (2005).

\bibitem{ManRichShapeFM} M. D. Kuz'min, M. Richter, and A. N. Yaresko,
Phys. Rev. B \textbf{73}, 100401(R) (2006).

\bibitem{bcc4}D. W. Wood and  N. W. Dalton,
Phys. Rev. \textbf{159}, 384 (1967).

\bibitem{Lovesey1977} S.W. Lovesey,
J. Phys. C: Solid State Phys. {\bf 10}, L455
 (1977). 

\bibitem{Sun2006} Shih-Jye Sun and Hsiu-Hau Lin,
Eur. Phys. J. B {\bf 49}, 403 (2006).


\bibitem{Katanin2} A.N. Ignatenko, A. A. Katanin, and  V. Yu. Irkhin,
JETP Lett. \textbf{97}, 209 (2013).

\bibitem{exp1}
F. J. Lazaro, J. Bartolome, R. Burriel, J. Pons, J. Casabo, and P. R.
Nugteren,
J. de Physique {\bf C8}, 825 (1988).

\bibitem{exp2}
 G. Cao, S. Chikara, E. Elhami, X.N. Lin, and P. Schlottmann, Phys. Rev. B
 {\bf 71}, 035104 (2005).

\bibitem{exp3}Lei Zhang, Jiyu Fan, Li Li, Renwen Li, Langsheng Ling,
Zhe Qu, Wei
Tong, Shun Tan, and Yuheng Zhang,
Europhys. Lett. {\bf 91}, 57001 (2010).


\end{thebibliography}
\end{document}